\documentclass[11pt]{article}
\usepackage{authblk}
\usepackage[utf8]{inputenc}
\usepackage{graphicx,xcolor}
\usepackage{amsmath}
\title{The $^{29}$F nucleus as a lighthouse on the coast of the island of inversion}
 
\author[1,2,$\star$]{L. Fortunato}
\author[1,2]{J. Casal}
\author[3]{W. Horiuchi}
\author[4]{Jagjit Singh}
\author[1,2]{A. Vitturi}
\affil[1 ]{\small Dip. Fisica e Astronomia ``G. Galilei'', Univ. Padova, via Marzolo 8, I-35131, Padova, Italy}
\affil[2 ]{\small I.N.F.N.~Sez. di Padova, via Marzolo 8, I-35131, Padova, Italy}
\affil[3 ]{\small Department of Physics, Hokkaido University, Sapporo 060-0810, Japan}
\affil[4 ]{\small Research Center for Nuclear Physics (RCNP), Osaka University, Ibaraki 567-0047, Japan}
\affil[$\star$ ]{\small Corresponding author: fortunat@pd.infn.it}
\date{July 2020}

\begin{document}

\maketitle
\section*{Preface}
The exotic, neutron-rich and weakly-bound isotope $^{29}$F stands
out as a waymarker on the southern shore of the island of inversion, a
portion of the nuclear chart where the effects of nuclear forces
lead to a reshuffling of the single particle levels and to a
reorganization of the nuclear structure far from stability.
This nucleus has become very popular, as new measurements allow
to refine theoretical models. We review the latest developments and
suggest how to further assess the structure by proposing predictions
on electromagnetic transitions that new experiments of Relativistic
Coulomb Excitation should soon become able to measure.

\section*{Introduction}
\begin{figure}
\begin{center}
      \includegraphics[width=0.8\linewidth]{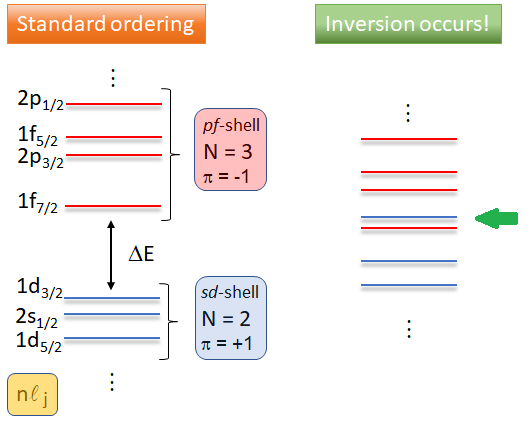}
\end{center}
    \caption{{\bf Standard ordering of shell model energy levels and typical inversion mechanism.} The $N=2$ {\it sd-}shell and the $N=3$ {\it pf-}shell with positive and negative parity $\pi$ respectively are shown on the left in the standard ordering (states are labeled by the standard set of quantum number $n\ell_j$). Inversion occur (right) when the shell gap, $\Delta E$, associated with the filling of 20 neutrons, disappears and one level (or more) of the $N=3$ {\it pf-}shell gets lower than one (or more) of the levels of the $N=2$ {\it sd-}shell.}
    \label{fig:inv}
\end{figure}

Weakly bound nuclear systems severely test our ability to disentangle the mysteries and oddities of the nuclear interactions and how nuclear systems achieve stability. One of the latest conundrums in this respect is the structure of the exotic 29-fluorine isotope. With 9 protons and 20 neutrons, it is located almost on the edge of the stability valley of the nuclide chart, very close to the neutron drip-line, 
i.e.~the dividing line ($S_\textrm{n}=0$, null separation energy) between bound and unbound neutron-rich nuclei.  Pioneering researches on this isotope have recently skyrocketed, as more and more theoretical and experimental papers are being published.  The main interest lies in understanding if it belongs to the so-called island of inversion, a portion of the nuclear chart where the standard list of single-particle energy levels (that are filled by nucleons, very similarly to the atomic physics counterpart, in a sort of {\it Aufbauprinzip}, a.k.a. the building-up principle, i.e. the rule that states how electronic orbitals are filled up in the atomic shell model) shows an inversion between orbitals belonging to the {\it sd-} and {\it pf-}shells, see Fig.~\ref{fig:inv}.
This is crucial to ascertain the presence of a neutron halo (a diffused tail of nuclear matter that spreads around the central core) and its extent.

This problem relates also to the disappearance of the $N=20$ neutron magic number when moving away from doubly-magic nuclei. This shell evolution has profound connections with tensor nuclear interactions \cite{Ots, Ots2} and with the deformation of the nuclear surface, occurring typically at mid-shell.
This system, due to its mass $A=29$, is still a hard nut to crack for {\it ab initio} approaches exploiting gargantuan numerical calculations. At present, simpler schematic models based on a few degrees of freedom are the preferred way to discern any pattern in the limited set of spectroscopic data.

\section*{Theoretical approaches and new experiments}
In one of the earliest theoretical three-body models for this nucleus \cite{noi}, the interactions between the 27-fluorine core and each of the two neutrons were tuned to reproduce the scarcely known properties of the binary subsystem 28-fluorine. Four different possible scenarios have been envisaged, called A, B, C, D, that we might rename as standard, intruder, degenerate and inverted scenarios, referring to the order of single particle levels and resonances appearing in the phase-shifts. The consequences of these choices have been analyzed and the relationships between the matter radius and the two-neutron separation energy with the percentage of $p$-wave content of the ground-state wave function have been explored \cite{noi}. These results show a mild to moderate sensitivity of these properties on the composition of the ground-state configuration and hint at the presence of a moderate halo structure connected with the $p$-orbital. 
The main reason to set up such a model is that $^{28}$F is neutron unbound with ground state at 220(50) keV and an unassigned resonance at 810 keV according to Ref. \cite{Chr1,Chr2}, where the g.s. value has now been  re-measured in Ref. \cite{Rev} to  199(6) keV, and the correlations between the two neutrons bring the nuclear system $^{29}$F back to stability, in a way similar to $^6$He and $^{11}$Li. When the binary subsystems are all unbound, as in all these cases mentioned, we speak of a Borromean nucleus, the archetypal system for studying two-neutron correlations. 
Another recent theoretical work \cite{Mic}, based on the Gamow shell model that treats bound and continuum states on the same footing, was applied to neutron-rich fluorine isotopes up to $^{31}$F. The comparative plots of one-body density distributions suggest that $^{31}$F has a halo structure with neutrons occupying an orbital with mixed configurations: 2$p_{3/2}$, 1$d_{3/2}$ and 1$f_{7/2}$, while $^{27}$F and $^{29}$F have instead a smaller radius, less compatible with a halo. These calculations produce a $^{29}$F wave function where the $p_{3/2}$ components amount to only a few percent \cite{Mic2}, more akin to the standard scenario A above. 
A novel mechanism to form the halo structure in $^{31}\text{F}$ was discussed in Ref. \cite{Mas}, in which the effect of pairing on the 
energy gaps between these orbits are essential
to determine the configuration mixing.
This puzzle is not to be resolved by theory alone, without further input from experiments.

Now, a new proton/neutron removal experiment on exotic beams \cite{Rev}, performed at the Radioactive Isotope Beam Factory (RIBF) of the RIKEN Nishina Center in Japan, indicates that $^{28}$F exists as a resonant state at about 199 keV above the $^{27}\text{F}+\textrm{n}$ threshold with a dominant $\ell=1$ orbital angular momentum component (79\%). This measurement is consistent with the position of the $p$-wave resonance in models C and D above, i.e.~200 keV (cfr.~Table 1 in Ref. \cite{noi}, using some prescriptions from Ref. \cite{Hor}). The analysis of this measurement is in stark contrast with the conclusions of Ref. \cite{Chr1} where it is stated that ``\dots $pf$ shell intruder components play only a small role in the ground state structure of $^{28}$F\dots''. 
The new data are interesting because they also show the presence of a resonance at about 0.9 MeV with a strong $\ell=2$ component (72\%). This is located half-way between scenarios C and D. 

\begin{figure}\centering
    \includegraphics[width=0.5\linewidth]{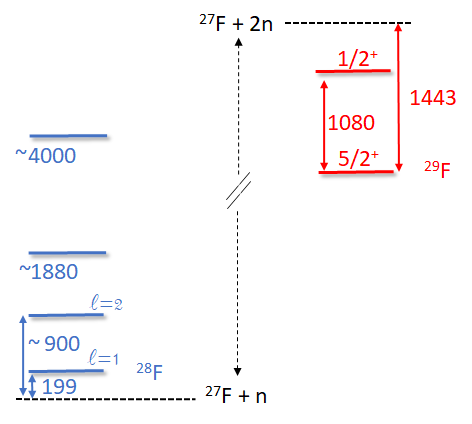}
    \caption{{\bf Synopsis of known experimental data on $^{28,29}$F}. All energies are in keV (not to scale) from Refs. \protect\cite{Rev,Chr1,Chr2,Door}. States in red are labelled by the $J^\pi$ quantum numbers and energies are referred to the $^{27}$F+2n threshold.
    States in blue are inferred from the $^{29}\text{F}(-1\textrm{n})$ column of Fig.~2 of Ref.~\protect\cite{Rev}, and correspond only to the states decaying to the ground state~of $^{27}$F. They are labelled by the orbital angular momentum quantum number, $\ell$, when available. Energies are referred to the $^{27}$F+n threshold.} 
    \label{fig:data}
\end{figure}

Previous measurements \cite{Door} found a resonance at 1080(18) keV in $^{29}$F, to which the $J^\pi=1/2^+$ quantum numbers were assigned. The analysis of shell model calculations with effective interactions, contained in the same paper, suggested that the ground state is probably a $5/2^+$ state, arising from the unpaired proton in the 1$d_{5/2}$ orbital, while the $1/2^+$ state arises mainly from the coupling of this proton with the 2$^+$ excitation of $^{28}$O. The latter is still unmeasured and requires further confirmation. The fact that the two-neutron separation energy was measured by Gaudefroy et al. \cite{Gau} at 1443 (436) keV, implies that this excited state is also bound, a singular coincidence for such a neutron-rich system that awaits further confirmation through the observation of gamma-ray transitions to the $^{29}$F ground state. The neutron-neutron correlations at $N=20$ are probably so strong as to make room not just for one, but for two bound states!
These data are collected in Fig. \ref{fig:data} for the purposes of the present perspective, but we refer the reader to the original papers for more details.

When the older data \cite{Door,Gau} on $^{29}$F were published, a theoretical study \cite{Mac} interpreted them as arising from the Nilsson deformed harmonic oscillator {\it sd}-states ($N=2$ shell) within the particle-rotor model with a moderate deformation. Now, the newly measured dominant $\ell=1$ component is incompatible with this view, apparently the model space should be enlarged to include intruder configurations from the $N=3$ shell.
Early indications that a large mixing with intruder 2p-2h and 4p-4h configuration is indeed necessary to help stabilizing the heavier fluorine isotopes came from Monte Carlo shell model calculations with the SDPF-M interaction \cite{Fal}.

\subsection*{New results}
\begin{figure}\centering
    \includegraphics[width=0.6\linewidth]{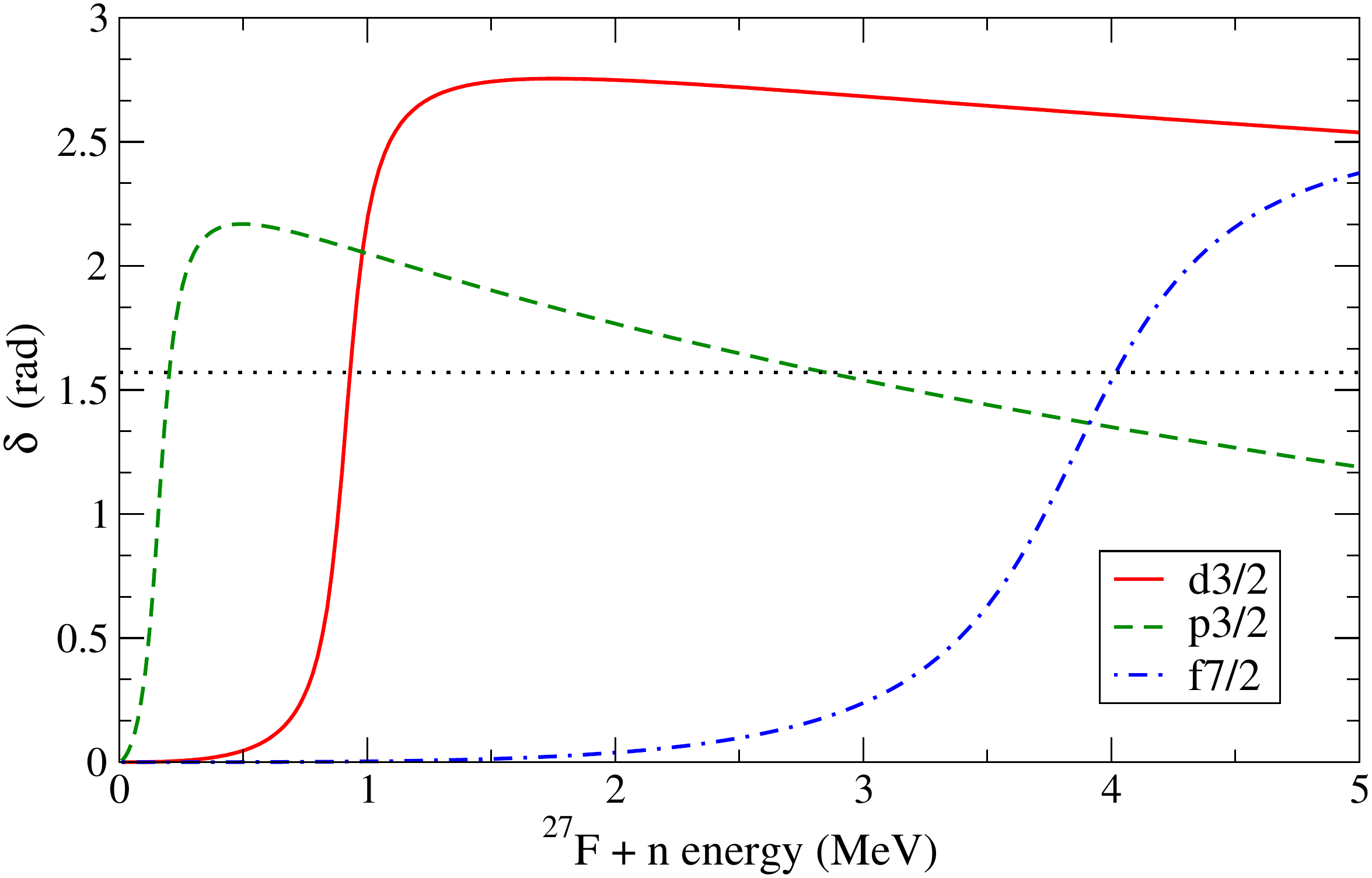}
    \caption{{\bf Phase-shifts, $\delta$, for the $^{27}\text{F}+\textrm{n}$ system in the D$^\flat$ scenario} as a function of the neutron-core relative energy. Adjusting the red curve to reproduce the $d$-resonance at about 0.9 MeV, also gives the $f$-wave state (blue curve) at about 4 MeV. }
    \label{fig:phase}
\end{figure}

\begin{figure}\centering
 \includegraphics[width=1.0\linewidth]{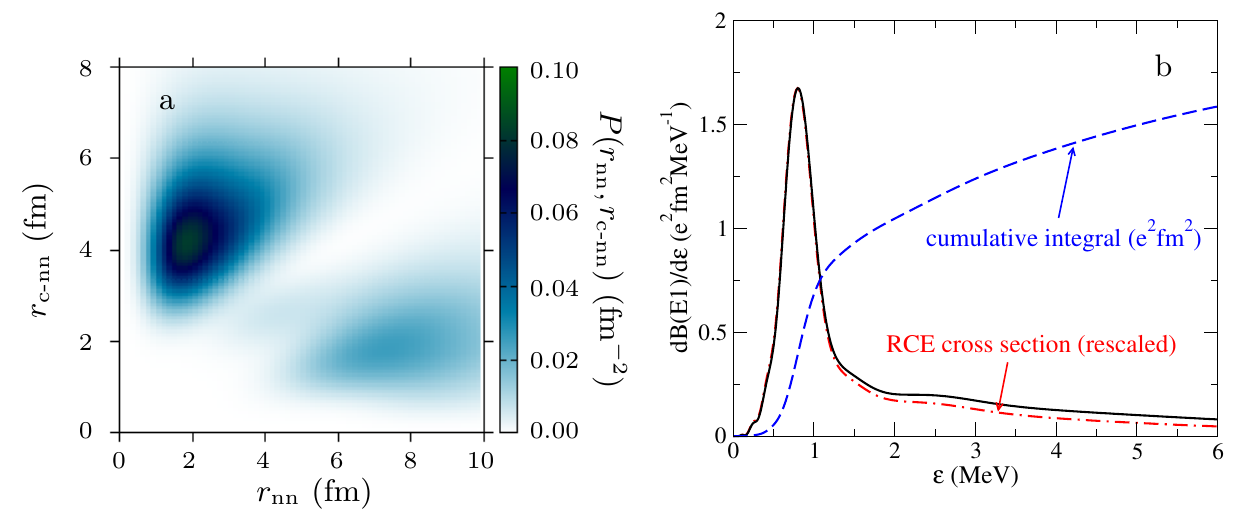}
    \caption{
    {\bf Results on $^{29}$F within the new $D^\flat$ scenario}.
    {\bf a.} Ground-state probability density of $^{29}$F as a function of the distance between the two valence neutrons ($r_\textrm{nn}$) and that between their center of mass and the core ($r_\textrm{c-nn}$). The maximum probability density corresponds to the dineutron configuration. {\bf b.}  Electric dipole ($E1$) strength function from the ground state to continuum as a function of the $^{27}\text{F}+\textrm{n}+\textrm{n}$ energy. The dashed line indicates the cumulative integral. The dash-dotted line is the corresponding Relativistic Coulomb Excitation (RCE) cross section, scaled to the same maximum to illustrate the decreasing proportionality with the energy.}
    \label{fig:29F}
\end{figure}

In view of the recent experimental results, we have adjusted our model for $^{28}$F to the new data, producing the scenario D$^\flat$, intermediate between C and D of Fig.~1 of Ref.~\cite{noi}. By simply adjusting the position of the $d$-wave resonant state, we get also an $f$-wave resonance at about the same energy of one of the states indicated in Fig.~2 of Ref.~\cite{Rev}. The phase-shifts are shown in Fig.~\ref{fig:phase} and the energy at which they cross 90$^\circ$ is the position of the resonant state. Using this updated model, we have performed new three-body ($^{27}\text{F}+\textrm{n}+\textrm{n}$) calculations for $^{29}$F to investigate its ground-state properties (as in Ref.~\cite{noi}) and, for the first time, the effect of configuration mixing on its dipole response. We obtain, as expected due to parity inversion, a ground state dominated by a $(2p_{3/2})^2$ component of 57.5\%, followed by 28.1\% of $(1d_{3/2})^2$ and a smaller 6\% of $(1f_{7/2})^2$. The dominant $p$-wave configuration leads to a non-negligible increase of the matter radius with respect to the $^{27}$F core, with $\Delta R = 0.192$ fm, which may be an indication of moderate halo formation in $^{29}$F. Intruder components seem to favor also the dineutron configuration, which becomes more pronounced in the wave-function density as a direct consequence of the mixing. This is shown in Fig.~\ref{fig:29F}a. 

From the ground state, we have considered electric dipole transitions into the continuum. It is known that two-neutron halo nuclei, and in general weakly-bound nuclei at the driplines, exhibit large $E1$ transition probabilities (and therefore large cross sections in Coulomb excitation experiments)~\cite{Aum, Nak, AN}. Since the halo structure is favored by $p$-wave intruder components, a measurement of the $E1$ strength could confirm whether $^{29}$F belongs to the island of inversion. Another way to pin down the content of p-waves is through neutron knock-out reactions. Our prediction using the three-body model constrained by the new data on $^{28}$F is shown in Fig.~\ref{fig:29F}b. With the present model, the $B(E1)$ distribution is strongly concentrated at low $^{27}\text{F}+n+n$ energies. The integrated $E1$ strength up to 6 MeV is 1.59 e$^2$fm$^2$, which is a large fraction of the total cluster sum rule, that amounts to 1.92 e$^2$fm$^2$. Note the presence of a prominent peak at about 0.8 MeV above the three-body threshold, which exhausts about 50\% of the total strength. This peak arises from a resonant state, which has been identified as an eigenstate of a resonance operator using the method proposed in Ref. \cite{cas19}. The composition of this state yields 73\% of $(2p_{3/2})(1d_{3/2})$ components, so that the computed $E1$ distributions arises mainly from $(2p_{3/2})^2\rightarrow (2p_{3/2})(1d_{3/2})$ and $(1d_{3/2})^2\rightarrow (2p_{3/2})(1d_{3/2})$ transitions.

It is worth noting that the $E1$ probability is highly sensitive to the ground-state radius and configuration mixing. In particular, the same calculations using the scenario A (standard shell-model order) of Ref.~\cite{noi} yields a reduction of the total $B(E1)$ by $\sim 40$\%. Such a difference should lead to very different experimental curves, making this observable a good probe for the configuration mixing and parity inversion in this region of the nuclear chart. 

\subsection*{Estimate of Relativistic Coulomb Excitation cross-section}
As a final note, we can estimate the cross-section for Relativistic Coulomb Excitation (RCE) to the continuum, which in the semi-classical approach it is just proportional to the distribution of electromagnetic strength \cite{Win,Ber}, i.e.~$$\frac{d\sigma}{d\varepsilon} =(\alpha Z_\textrm{2})^2 \sum_{\pi\lambda\mu} \Bigl(\frac{\Delta E}{\hbar c}\Bigr)^{2\lambda-2}\frac{dB}{d\varepsilon}(\pi\lambda; \textrm{gs}\rightarrow \textrm{Cont.})\mid G_{\pi\lambda\mu}(\beta^{-1})\mid^2g_\mu(\xi)$$ 
where $\alpha$ is the fine structure constant, $\Delta E= S_\textrm{2n}+\epsilon$, $G_{\pi\lambda\mu}$ are the relativistic Winther-Adler functions and $g$ is a function of the adiabaticity parameter $\xi$.
Note that the proportionality factor, despite having no direct dependence on energy when $\lambda=1$, is not a constant, because the function $g$ has a decreasing dependence on energy through $\xi$ (at this bombarding energy it amounts to a factor of 2 over the first 10 MeV's). This modulates slightly the dipole distribution, by flattening its tail (see Fig.~\ref{fig:29F} right panel).
Choosing the same kinematic parameters of the experiment at RIBF \cite{Rev}, i.e. a beam of $^{29}$F ($Z_\textrm{1}=9$) at an energy of $E_\textrm{B}=235$ MeV/nucleon, impinging on a target of $^{208}$Pb ($Z_\textrm{2}=82$), we get an estimate for the total cross-section of about 600 mb, integrated up to 15 MeV (with more than 90\% of the cross-section concentrated below 5-6 MeV). As discussed in Ref.~\protect\cite{Hor}, the nuclear contribution is not negligible,
about half of the Coulomb contribution, in case of $^{31}\text{Ne}$.
With a simple geometric model, the nuclear contribution can
roughly be estimated as $2\pi \sqrt{\frac{5}{3}}(R(^{27}\text{F})+R(^{208}\text{Pb}))\Delta R$, resulting in $\approx 200$ mb.
The semiclassical theory allows also to estimate the forward-peaked angular range for this process to about $\Delta\theta \sim 2Z_\textrm{1}Z_\textrm{2}e^2/(R A_\textrm{1} E_\textrm{B}) \sim 0.22^\circ$ \cite{Win}. Our predictions may guide the interpretation of future experiments focused on the structure of $^{29}$F and the extent of the island of inversion.
Let us also mention that a new method, alternative to the Equivalent Photon Method, has recently been proposed \cite{Mor} to extract $B(E1)$ strengths, disentangling nuclear and coulomb contributions.

\subsection*{Note added during revision}
Slightly after our Perspective paper has been submitted, another paper was published in Physical Review Letters \cite{Bag20} where new measurements of reaction cross-sections are reported for $^{27,29}$F on C at RIKEN. From the large value of the cross-section, out of the systematic trend for fluorine isotopes, the authors conclude that $^{29}$F is the heaviest Borromean two-neutron halo nucleus to date and explain their findings with a large occupancy of the $p_{3/2}$ orbital, supported by shell-model calculations and compared with coupled-cluster calculations. This finding is fully consistent with our discussion in Table III of Ref. \cite{noi}, despite the suggested value of $\Delta R =0.35$ fm being somewhat larger than our predicted value for $\Delta R$, thus pointing to a larger halo structure. The comparison of our three-body calculations with the shell-model results presented in Ref. \cite{Bag20} shows a reasonable agreement.

\section*{Outlook}
In conclusion, we believe that new experiments by Revel and collaborators \cite{Rev} and by Bagchi and collaborators \cite{Bag20} are clearing the field of any doubt on the theoretical interpretation on the structure of the interesting $^{29}$F nucleus. Isotopes sitting near the southern shore of the island of inversion are crucial for our understanding of nuclear stability in general and for the peculiar interplay of nuclear forces that determines the insurgence of a small $p$-wave halo, the shell-evolution and the disappearance of the $N=20$ magic number in this region of the nuclear chart. We propose that Relativistic Coulomb Excitation experiments might lead to additional confirmations of these new findings.

\section*{Aknowledgements}
This work was supported by SID funds 2019 (Università degli Studi di Padova, Italy) under project No.~CASA\_SID19\_01, by JSPS KAKENHI Grants No.~18K03635, No.~18H04569, and No.~19H05140 and by the collaborative research programs 2020, Information Initiative Center, Hokkaido University. 

    



\begin{thebibliography}{xxxx}
\bibitem{Ots} Otsuka, T. et al., Magic Numbers in Exotic Nuclei and Spin-Isospin Properties of the NN Interaction. Phys. Rev. Lett. {\bf 87,} 082502 (2001).
\bibitem{Ots2} Otsuka, T., Suzuki, T., Fujimoto, R., Grawe, R.H. \&  Akaishi, Y. Evolution of Nuclear Shells due to the Tensor Force. Phys. Rev. Lett. {\bf 95,} 232502 (2005).
\bibitem{noi} Singh, J. Casal, J., Horiuchi, W., Fortunato, L. \& Vitturi, A. Exploring two-neutron halo formation in the ground state of $^{29}$F within a three-body model. Phys. Rev C {\bf 101,} 024310 (2020).
\bibitem{Mic} Michel, N., Li, J.G., Xu, F.R., \& Zuo, W.  Two-neutron halo structure of $^{31}$F. Phys. Rev. C {\bf 101,} 031301 (R)  (2020).
\bibitem{Mic2} Michel, N. private communication (March 2020).
\bibitem{Mas} Masui, H., Horiuchi, W., \&  Kimura, K. Two-neutron halo structure of $^{31}$F and a novel pairing antihalo effect. Phys. Rev. C {\bf 101,} 041303(R) (2020).
\bibitem{Rev} Revel A. et al., Extending the Southern Shore of the Island of Inversion to $^{28}$F. Phys. Rev. Lett. {\bf 124,} 152502 (2020).
\bibitem{Hor} Horiuchi, W., Suzuki, Y., Capel, P. \& Baye, D. Probing the weakly-bound neutron orbit of $^{31}$Ne with total reaction and one-neutron removal cross sections. Phys. Rev. C {\bf 81,} 024606 (2010).	
\bibitem{Chr1} Christian, G. et al., Exploring the Low-$Z$ Shore of the Island of Inversion at $N=19$. Phys. Rev. Lett. {\bf 108,} 032501 (2012).
\bibitem{Chr2} Christian, G. et al., Spectroscopy of neutron-unbound $^{27,28}$F. Phys. Rev. C {\bf 85,} 034327 (2012).
\bibitem{Door} Doornenbal, P. et al., Low-$Z$ shore of the “island of inversion” and the reduced neutron magicity toward $^{28}$O. Phys. Rev. C {\bf 95,} 041301(R) (2017).
\bibitem{Gau} Gaudefroy L. et al., Direct Mass Measurements of $^{19}$B, $^{22}$C, $^{29}$F, $^{31}$Ne, $^{34}$Na and Other Light Exotic Nuclei. Phys. Rev. Lett. {\bf 109,} 202503 (2012).
\bibitem{Mac} Macchiavelli, A.O. et al., Structure of $^{29}$F in the rotation-aligned coupling scheme of the particle-rotor model. Phys. Lett. B {\bf 775,} 160-162 (2017).
\bibitem{Fal} Fallon P. et al., Two-proton knockout from $^{32}$Mg: Intruder amplitudes in $^{30}$Ne and implications for the binding of $^{29,31}$F. Phys. Rev. C {\bf 81,} 041302(R) (2010).
\bibitem{Aum} Aumann, T. Low-energy dipole response of exotic nuclei. Eur. Phys. J. A {\bf 55}, 234 (2019).
\bibitem{cas19} Casal, J. \& Gómez-Camacho, J. Identifying structures in the continuum: Application to $^{16}$Be. Phys. Rev. C {\bf 99,} 014604 (2019).
\bibitem{Win} Winther, A. \& Alder, K. Relativistic Coulomb excitation. Nucl. Phys. A {\bf 319,} 518-532 (1979).
\bibitem{Ber} Bertulani, C. \& Baur, G. Electromagnetic processes in relativistic heavy ion collisions. Phys. Rep. {\bf 163,} no. 5 \& 6, 299-408 (1988).
\bibitem{Nak} Nakamura, T. et al. Observation of strong low-lying E1 strength in the two-neutron halo 
nucleus $^{11}$Li, Phys. Rev. Lett. {\bf 96,} 252502 (2006).
\bibitem{AN} Aumann, T. \&  Nakamura, T. The electric dipole response of exotic nuclei, Phys. Scr. {\bf 2013,} 014012 (2013).
\bibitem{Mor} Moro, A.M., Lay, J.A., G\'omez-Camacho, J., Determining B(E1) distributions of weakly bound nuclei from breakup cross sections using Continuum Discretized Coupled Channels calculations. Application to 11Be. Preprint at  https://arxiv.org/abs/2004.14612 (2020).
\bibitem{Bag20} Bagchi, S.  et al., Two-neutron halo is unveiled in $^{29}$F, Phys. Rev. Lett. {\bf 124,} 222504 (2020). 
\end{thebibliography}
\end{document}